\documentclass[12pt,preprint,twocolappendix,numberedappendix]{emulateapj}

\usepackage{subeqnarray}
\usepackage{amsmath}
\usepackage{hyperref}
\usepackage{ulem}
\usepackage{stmaryrd}
\hypersetup{colorlinks=true,linkcolor=black,citecolor=blue,urlcolor=black}
\usepackage{lineno}
\usepackage{amssymb}
\usepackage{pifont}
\usepackage{verbatim}
\usepackage{natbib}
\usepackage[counterclockwise, figuresleft]{rotating}

\def\tt{\bf   }

\newcommand{\Eq}[1]{Equation\,(\ref{#1})}

\newcommand{\Sec}[1]{Section~\ref{#1}}

\newcommand{\Fig}[1]{Figure~\ref{#1}}

\newcommand{\Tab}[1]{Table \ref{#1}}

\newcommand {\exocam} {{\ttfamily ExoCAM}}

\setcitestyle{citesep={,}}
\begin{document}

\slugcomment{Accepted at ApJ}
\shorttitle{Hurricane genesis on terrestrial exoplanets}
\shortauthors{Komacek, Chavas, \& Abbot}

\title{Hurricane genesis is favorable on terrestrial exoplanets orbiting late-type M dwarf stars}
\author{Thaddeus D. Komacek$^{1}$, Daniel R. Chavas$^{2}$, and Dorian S. Abbot$^{1}$} \affil{$^{1}$Department of the Geophysical Sciences, The University of Chicago, Chicago, IL, 60637, USA \\ $^{2}$Department of Earth, Atmospheric, and Planetary Sciences, Purdue University, West Lafayette, IN, 47907, USA  \\
\url{tkomacek@uchicago.edu}} 
\begin{abstract}
Hurricanes are one of the most extreme storm systems that occur on Earth, characterized by strong rainfall and fast winds. The terrestrial exoplanets that will be characterized with future infrared space telescopes orbit M dwarf stars. As a result, the best observable terrestrial exoplanets have vastly different climates than Earth, with a large dayside-to-nightside irradiation contrast and relatively slow rotation. Hurricanes may affect future observations of terrestrial exoplanets because they enhance the vertical transport of water vapor and could influence ocean heat transport. In this work, we explore how the environment of terrestrial exoplanets orbiting M dwarf stars affects the favorability of hurricane genesis (formation). To do so, we apply metrics developed to understand hurricane genesis on Earth to three-dimensional climate models of ocean-covered exoplanets orbiting M dwarf stars. We find that hurricane genesis is most favorable on intermediate-rotating tidally locked terrestrial exoplanets with rotation periods of $\sim 8-10~\mathrm{days}$. As a result, hurricane genesis is most favorable for terrestrial exoplanets in the habitable zones of late-type M dwarf stars. The peak in the favorability of hurricane genesis at intermediate rotation occurs because sufficient spin is required for hurricane genesis, but the vertical wind shear on fast-rotating terrestrial exoplanets disrupts hurricane genesis. We find that hurricane genesis is less favorable on slowly rotating terrestrial exoplanets, which agrees with previous work. Future work using simulations that resolve hurricane genesis and evolution can test our expectations for how the environment affects the favorability of hurricane genesis on tidally locked terrestrial exoplanets. 
\end{abstract}
\section{Introduction}
\indent Hurricanes are tropical cyclones that have intensified such that their maximum wind speeds are $\ge 33~\mathrm{m}~\mathrm{s}^{-1}$. Hurricanes are an integral part of Earth's climate, as $\sim 100$ tropical cyclones form on Earth per year \citep{Emanuel:2003aa,Hoogewind:2020aa} with the systems that intensify into hurricanes leading to heavy rainfall and strong winds. Hurricanes also play a role in the global-mean climate of a planet, as they affect oceanic heat transport and mixing \citep{Emanuel:2001aa,Sriver:2007aa,Jansen:2009aa}. Notably, \cite{Jansen:2009aa} found that hurricanes on modern Earth likely reduce the oceanic equator-to-pole heat transport, but if hurricanes could form closer to the equator than they do on Earth they would enhance the equator-to-pole heat transport.  As a result, determining the conditions that are necessary for hurricane genesis is crucial for understanding planetary climate.  \\
\indent Studies of hurricanes on Earth have shown that empirical metrics can be used to determine the environmental favorability of hurricane genesis \citep{Emanuel:2010aa,Tang:2012aa,Camargo:2014,Vecchi:2019}. These studies have found that on Earth, larger absolute vorticity (spin) and hotter sea surface temperatures favor hurricane genesis. Notably, ventilation of hurricanes by import of dry, low-entropy air in the middle troposphere acts to disrupt the energetics of hurricanes \citep{Tang:2010aa, Tang:2012a, Riemer:2015aa, Chavas:2017aa}. As a result, the environment can resist hurricane activity through the process of ventilation \citep{Hoogewind:2020aa}. These environmental metrics are especially useful because they are intrinsically large-scale in nature and thus they can be resolved in coarse-grained climate models even if hurricanes themselves cannot be. A key unanswered question of hurricane genesis and evolution is how the ventilation of hurricanes depends on the global-mean planetary climate itself.    \\
\indent A variety of recent work has studied how hurricane activity varies with the thermodynamic climate forcing on Earth. These studies range from cold climates in the Last Glacial Maximum \citep{Korty:2012aa} to very warm climates in extreme greenhouse states \citep{Korty:2017aa}, as well as under future anthropogenic warming \citep{Vecchi:2019,Zhou:2019}. Additional work has also studied how hurricane activity varies in simplified Earth-like aquaplanet simulations with uniform and meridionally varying thermal forcing \citep{Merlis:2016,Walsh:2020}. Similarly, recent idealized modeling work has understood how hurricane activity depends fundamentally on dynamical climate forcing by varying planetary rotation rate and size relative to Earth values \citep{Chavas:2019aa}, as well as with a uniform Coriolis effect \citep{Reed:2015aa}.
Additionally, \cite{Cronin:2019aa} limited the availability of surface moisture, finding that hurricanes could still occur in dry climates. \\
\indent As discussed above, substantial research has demonstrated how the environmental favorability for hurricane genesis varies in Earth's climate system. In this work, we apply these techniques to understand potential hurricane activity on planets with very different climates than Earth. Planets orbiting M dwarf stars provide a novel climate regime to test theories for hurricane genesis, as they orbit close-in to their host star and as a result are likely tidally locked. Because their rotation period is set by their orbital semi-major axis, the atmospheres of planets in the habitable zones of M dwarfs can be in a broad range of dynamical regimes dependent on their host star type \citep{Haqq2018}. A wide range of previous work has studied the climate of temperate terrestrial planets orbiting M dwarf stars, finding that they have a diverse array of possible climates depending on planetary and host star properties \citep{Joshi:1997,Merlis:2010,Selsis:2011aa,Leconte:2013aa,Yang:2013,Hu:2014aa,Wang:2014,Yang:2014,Carone:2015aa,Koll:2015,Koll:2016,Turbet:2016aa,Wang:2016,Boutle:2017aa,Fujii:2017aa,kopparapu2017,Noda:2017aa,Wolf:2017aa,Bin:2018aa,Haqq2018,Lewis:2018aa,way:2018,Chen:2019aa,DelGenio:2019,Fauchez:2019aa,Komacek:2019aa,Wolf:2019aa,Yang:2019aa,Suissa:2020aa}. As a result, the dynamic and thermodynamic constraints on the environmental favorability of hurricane genesis will likely change greatly over the broad range of possible habitable terrestrial exoplanets orbiting late-type stars. \\
\indent To date, there have been two studies of hurricane genesis on terrestrial exoplanets orbiting M dwarf stars. \cite{Bin:2017aa} studied the environmental favorability of hurricane genesis by analyzing the genesis potential index \citep{Emanuel:2004aa,Emanuel:2010aa}, a metric for the favorability of hurricane genesis, from the general circulation model (GCM) simulations of slowly rotating aquaplanets of \cite{Wang:2016}. \cite{Bin:2017aa} found that hurricanes are unlikely to occur on slowly rotating planets with a 28 day orbital period orbiting an M dwarf star with an effective temperature of 3700 K, but did not extend their simulations to the fast-rotating dynamical regime of planets orbiting late-type M dwarf stars. More recently, \cite{Yang:2020} performed GCM simulations with a horizontal resolution of $\approx 50$ km in order to directly simulate hurricanes on terrestrial planets orbiting M dwarf stars. \cite{Yang:2020} found that hurricanes can occur on tidally locked planets, provided that the sea surface temperature is warm enough and the background gas is not hydrogen and/or helium. Additionally, \cite{Yang:2020} found that hurricanes most commonly form in regions north and south of the substellar point, in agreement with the regions of highest environmental favorability found by \cite{Bin:2017aa}. However, these simulations used a sea surface temperature distribution that was fixed in time and only considered a limited set of planetary parameters. \\
\indent In this work, we improve on previous work that studied hurricanes on tidally locked terrestrial exoplanets by determining how the favorability of hurricane genesis depends on a broad range of possible planetary parameters. To do so, we apply established metrics for the environmental favorability of hurricane genesis to GCM simulations of planets with varying host star type, rotation period, incident stellar flux, planetary radius, surface gravity, and surface pressure. We find that the environmental favorability for hurricane genesis is strongly dependent on planetary parameters and host star type. We also show that hurricanes are more likely to occur on planets in the habitable zones of late-type M dwarfs than those of earlier-type M dwarf stars. This paper is organized as follows. In \Sec{sec:methods}, we describe the setup of our GCM simulations, and in \Sec{sec:metrics} we derive the metrics we use to calculate the favorability of hurricane genesis. We present our results for the favorability of hurricane genesis as a function of planetary parameters in \Sec{sec:results} in order to determine which terrestrial exoplanets orbiting M dwarf stars are most favorable for hurricane genesis. We discuss how hurricanes may affect observable properties of terrestrial exoplanets in \Sec{sec:disc}, and conclude in \Sec{sec:conc}.  
\section{GCM setup}
\label{sec:methods}
To determine the favorability of hurricane genesis on terrestrial exoplanets orbiting M dwarf stars, we use the \exocam\footnote{\url{https://github.com/storyofthewolf/ExoCAM}}~GCM. \exocam~has been used in a wide range of studies of both early Earth and of terrestrial exoplanets \citep{Wolf:2015,Kopp:2016,kopparapu2017,Wolf:2017,Wolf:2017aa,Haqq2018,Komacek:2019aa,Komacek:2019ab,Suissa:2019aa,Wolf:2019aa,Yang:2019aa,Komacek:2020aa,Suissa:2020aa}, and is an upgraded version of the Community Atmosphere Model version 4 with a novel correlated-k radiative transfer scheme ({\ttfamily ExoRT}\footnote{\url{https://github.com/storyofthewolf/ExoRT}}) that can model terrestrial exoplanets near the inner edge of the habitable zone. In this work, we study how the favorability of hurricane genesis depends on a wide range of planetary parameters, including rotation period, incident stellar flux, planetary radius, surface gravity, and surface pressure.
This set of GCM simulations builds upon those presented in \cite{Komacek:2019aa,Komacek:2019ab}, and \cite{Komacek:2020aa}. This includes simulations for both intermediate and slow rotators with varying radius, gravity, and surface pressure. The simulations varying individual planetary parameters are for planets orbiting an M dwarf star with $T_\mathrm{eff} = 2600~\mathrm{K}$ and with a fixed incident stellar flux equal to that of Earth, without consistently varying the incident stellar spectrum with rotation period. However, we do consistently vary the incident stellar flux and rotation period together for planets orbiting M dwarf stars in separate model suites with $T_\mathrm{eff} = 2600~\mathrm{K},~3000~\mathrm{K},~3300~\mathrm{K}, ~\mathrm{and}~4000~\mathrm{K}$. The M dwarf host star spectra used in this work are taken from BT-SETTL models \citep{Allard:2007aa}. \\
\indent All of our simulations in this work consider terrestrial exoplanets that are tidally locked to their host star with zero obliquity and zero eccentricity and have an ocean that covers the entire planet. In the majority of the simulations in this work, we use a 50 m deep slab ocean and an atmosphere comprised purely of N$_2$ and H$_2$O as in \cite{kopparapu2017} and \cite{Komacek:2019aa}. However, we test the sensitivity of our results to these assumptions for the ocean depth and atmospheric composition by reducing the slab ocean depth to 1 m and by conducting additional GCM simulations including modern Earth-like abundances of CO$_2$ and CH$_4$. Our simulations all have the same horizontal resolution of $4^\circ \times 5^\circ$ and use 40 vertical levels. Our nominal timestep is 30 minutes, but for numerical stability we reduce the timestep to 7.5 minutes in our simulations of planets near the inner edge of the habitable zone. Simulations are run for at least 45 years and continued until they reach steady state in net top-of-atmosphere radiative flux. We use daily output (as in \citealp{Hoogewind:2020aa}) from the final year of simulation time to compute metrics for the environmental favorability of hurricane genesis. These metrics are defined in \Sec{sec:metrics}, and results of our analysis are presented in \Sec{sec:results}. For comparison, we also show the hurricane genesis favorability in the mean climate from a subset of simulations in Appendix \ref{sec:appa}. 
\section{Metrics for the favorability of hurricane genesis}
\label{sec:metrics}
\subsection{Combined environmental favorability metrics}
\begin{table}
\begin{center}
\begin{tabular}{| c | c | } 
\hline
{\bf Metric} & {\bf Threshold} \\
\hline
Ventilation Index (VI) & VI $\le 0.145$ \\
\hline
Lower-atmosphere Absolute Vorticity ($\eta$) & $\left|\eta\right| \ge 1.2 \times 10^{-5}~\mathrm{s}^{-1}$ \\
\hline
Maximum Potential Intensity ($u_p$) & $u_p \ge 33~\mathrm{m}~\mathrm{s}^{-1}$ \\
\hline
\end{tabular}
\caption{\textbf{We use environmental favorability thresholds for hurricane genesis that are based in the study of hurricanes on Earth.} Shown are the ventilation index, absolute vorticity, and maximum potential intensity thresholds we use to define hurricane genesis favorability.}
\label{table:params}
\end{center}
\end{table}
\indent In this work, we use a combination of both dynamic and thermodynamic metrics that are established in the study of the environmental favorability of hurricane genesis on Earth. Specifically, we use a combination of three metrics to define environmental favorability: the ventilation index (VI), absolute vorticity ($\eta$), and maximum potential intensity ($u_p$). The ventilation index captures the energetic effect of the import of low entropy (dry) air into the storm core by the environmental vertical shear.
Lower ventilation indices favor hurricane genesis and intensification. The absolute vorticity is used to determine if there is sufficient spin available for the generation of a cyclonic storm, while the maximum potential intensity characterizes the maximum possible wind speed of the storm for a given thermodynamic environment. \\
\indent \Tab{table:params} defines our environmental favorability thresholds for each metric, each of which is described in further detail below. We utilize a combination of the ventilation index and absolute vorticity
to determine if a hurricane is environmentally favorable --- specifically, the genesis threshold from both metrics must be satisfied at a given location on the planet. The environmental favorability thresholds for the ventilation index and absolute vorticity listed in \Tab{table:params} were defined by \cite{Hoogewind:2020aa} based on the 95th percentile of each metric for hurricane genesis on Earth, determined empirically from the late-20th century historical hurricane record. Favorability is calculated from output that is averaged over each day\footnote{In the remainder of this manuscript a ``day'' is equal to $86,400~\mathrm{s}$.} and is used to define the fraction of time that a region is favorable for hurricane genesis. We then use the local maximum potential intensity to determine if the maximum possible wind speed is fast enough for hurricane genesis to occur. \\
\indent The first environmental favorability metric we use is the ventilation index, defined by \cite{Tang:2012aa} as
\begin{equation}
\label{eq:vi}
\mathrm{VI} = \frac{u_{\mathrm{shr}}\chi_m}{u_p} \mathrm{,}
\end{equation}
where $u_\mathrm{shr}$ is the magnitude of the change in velocity across the troposphere (from a pressure of $\sigma = p/p_\mathrm{surf} = 0.85$ to the tropopause), $\chi_m$ is the middle-troposphere entropy deficit, calculated at $\sigma = 0.6$, and $u_p$ is the maximum potential intensity. We calculate the mid-troposphere entropy deficit as in \cite{Hoogewind:2020aa},
\begin{equation}
\chi_m = \frac{s^{sat}_m-s_m}{s^{sat}_{SST}-s_b} \mathrm{,}
\end{equation}
where $s^{sat}_m-s_m$ is the difference between the saturation entropy measured in the inner core of the hurricane ($s^{sat}_m$) and the environmental entropy ($s_m$), and $s^{sat}_{SST}-s_b$ is the difference between the saturation entropy at the sea surface temperature ($s^{sat}_{SST}$) and the entropy of the boundary layer ($s_b$). \cite{Hoogewind:2020aa} defined the entropy deficit using an annulus of fixed radii relative to each grid point; here we define it within each grid point directly to simplify the calculation. \\
\indent The maximum potential intensity is derived by considering the energetics of a hurricane, assuming that a mature hurricane can be modeled as an axisymmetric storm with an energy cycle that is similar to a Carnot cycle \citep{Emanuel:1986aa,Emanuel:2003aa}. \cite{Bister:2002aa} define the maximum potential intensity as
\begin{equation}
\label{eq:mpi}
u^2_p = \frac{T_s}{T_o}\frac{C_k}{C_D}\left[(\mathrm{CAPE}_s - \mathrm{CAPE}_b)\right]|_m \mathrm{,}
\end{equation}
where $T_s$ is the surface temperature, $T_o$ is the outflow temperature at the top of the hurricane, $C_k/C_d = 0.7$ is the ratio of the enthalpy exchange and drag coefficients, and $\left[(\mathrm{CAPE}_s - \mathrm{CAPE}_b)\right]|_m$ is the difference between the convective available potential energy (CAPE) of the sea-level air at saturation and the boundary layer air, evaluated at the radius of maximum wind. In this work, we use the algorithm of \cite{Hoogewind:2020aa} to calculate the ventilation index and maximum potential intensity, where the potential intensity calculation follows \cite{Bister:2002aa}. \\
\indent Additionally, we use a threshold for the absolute vorticity 
to determine the environmental favorability of hurricane genesis. The absolute vorticity is defined as \citep{Holton:2013}
\begin{equation}
\label{eq:vorticity}
\eta = \zeta + f = \hat{k} \cdot \left(\nabla \times {\bf u}\right) + 2 \Omega\mathrm{sin}\phi \mathrm{,}
\end{equation}
where $\zeta$ is the relative vorticity derived from the atmospheric circulation and $f$ is the planetary vorticity from the planetary rotation. In \Eq{eq:vorticity}, ${\bf u}$ is the horizontal wind speed, $\Omega$ is the planetary rotation rate, and $\phi$ is latitude. We calculate $\eta$ at $\sigma = 0.85$ using {\ttfamily windspharm}\footnote{\url{https://ajdawson.github.io/windspharm}} \citep{Dawson:2016}. 
\subsection{Other hurricane activity metrics}
\label{sec:othermetrics}
\indent In this work, we also consider two other hurricane activity metrics that we do not include as part of our combined metrics for the environmental favorability of hurricane genesis. The first metric is the ventilation-reduced maximum potential intensity, which we term $u_{p,\mathrm{VI}}$. $u_{p,\mathrm{VI}}$ is the maximum achievable wind speed for a hurricane in the presence of mid-level ventilation by dry, low-entropy air. The ventilation-reduced maximum potential intensity can be determined through the equilibrium solution for normalized intensity in the presence of ventilation, as in Equation (19) of \cite{Chavas:2017aa}:
\begin{equation}
\label{eq:normventvp}
\tilde{u}^3_{p,\mathrm{VI}} - \tilde{u}_{p,\mathrm{VI}} + \lambda = 0 \mathrm{.}
\end{equation}
In \Eq{eq:normventvp}, $\tilde{u}_{p,\mathrm{VI}} = u_{p,\mathrm{VI}}/u_p$ is the ratio of the ventilation-reduced maximum potential intensity and the maximum potential intensity. 
$\lambda$ is the normalized ventilation \citep{Tang:2010aa}, which is related to the ventilation index by a constant $c_1$ as
\begin{equation}
\lambda = c_1 \mathrm{VI} \mathrm{.}
\end{equation}  
Solving for $\tilde{u}_{p,\mathrm{VI}}$ in \Eq{eq:normventvp}, we find
\begin{equation}
\label{eq:vpvi}
\begin{aligned}
\tilde{u}_{p,\mathrm{VI}} = & \frac{\left(\sqrt{3} \sqrt{27\left(c_1 \mathrm{VI}\right)^2 - 4)} - 9c_1\mathrm{VI}\right)^{1/3}}{2^{1/3} 3^{2/3}} \\ & + \left(\frac{2}{3\sqrt{3} \sqrt{27\left(c_1 \mathrm{VI}\right)^2 - 4)} - 27c_1\mathrm{VI}}\right)^{1/3} \mathrm{.}
\end{aligned}
\end{equation} 
This solution is plotted in Figure 1 of \cite{Chavas:2017aa}: $\tilde{u}_{p,VI} = 1$ (i.e. no reduction in potential intensity) for very small ventilation index, and $\tilde{u}_{p,VI}$ decreases to a minimum of 0.577 (i.e., 42\% reduction) at the threshold value of ventilation; above this threshold the value jumps to zero as a hurricane cannot be supported at all. Note that each term in the general solution given by \Eq{eq:vpvi} has complex parts which cancel out, leaving a real solution for the relevant range of VI. \\
\indent We use the threshold VI for hurricane genesis of $\mathrm{VI}_\mathrm{thresh} = 0.145$ from \cite{Hoogewind:2020aa} to solve for $c_1$ as
\begin{equation}
\label{eq:c1}
c_1 = \frac{2}{3\sqrt{3} \mathrm{VI}_\mathrm{thresh}} \mathrm{.}
\end{equation} 
Given a local value for the ventilation index determined from \Eq{eq:vi}, we calculate the ventilation-reduced maximum potential intensity as the product of the maximum potential intensity from \Eq{eq:mpi} and the normalized ventilation-reduced maximum potential intensity from \Eq{eq:vpvi}. The ventilation-reduced maximum potential intensity is a convenient way to directly represent the detrimental effect of ventilation on the maximum potential intensity in a single metric that is grounded in existing theory. Note that this solution has yet to be explicitly applied to data for hurricanes on Earth even though its underlying components (potential intensity and ventilation index) have been extensively applied (e.g., \citealp{Tang:2012aa}). \\
\indent To compare our results to those of \cite{Bin:2017aa}, we also calculate the genesis potential index (GPI) \citep{Emanuel:2004aa,Emanuel:2010aa} from our GCM simulations. We calculate GPI as in Equation (1) of \cite{Bin:2017aa},
\begin{equation}
\label{eq:GPI}
\mathrm{GPI} = \left(10^5 \left|\eta\right|\right)^{3/2} \left(\frac{u_p}{70~\mathrm{m}\mathrm{s}^{-1}}\right)^3 \left(\frac{\mathrm{RH}}{50\%}\right)^3 \left(1 + 0.1u_\mathrm{shr}\right)^{-2} \mathrm{,}
\end{equation} 
where RH is the relative humidity measured in $\%$. As in the Earth-based analysis of \cite{Hoogewind:2020aa}, we find that GPI provides similar environmental favorability estimates to the combined metric used in this work (see Appendix \ref{app:gpi}). Note that \cite{Yang:2020} recently found that the environmental favorability estimated from GPI matches well with the locations of hurricane genesis in high-resolution simulations of terrestrial exoplanets. 
\section{The dependence of the favorability of hurricane genesis on planetary properties}
\label{sec:results}
\subsection{Combined environmental favorability metrics}
\subsubsection{Dependence on planetary parameters}
\begin{sidewaysfigure*}
\centering
\vspace{-3in}
\includegraphics[width=1\textwidth]{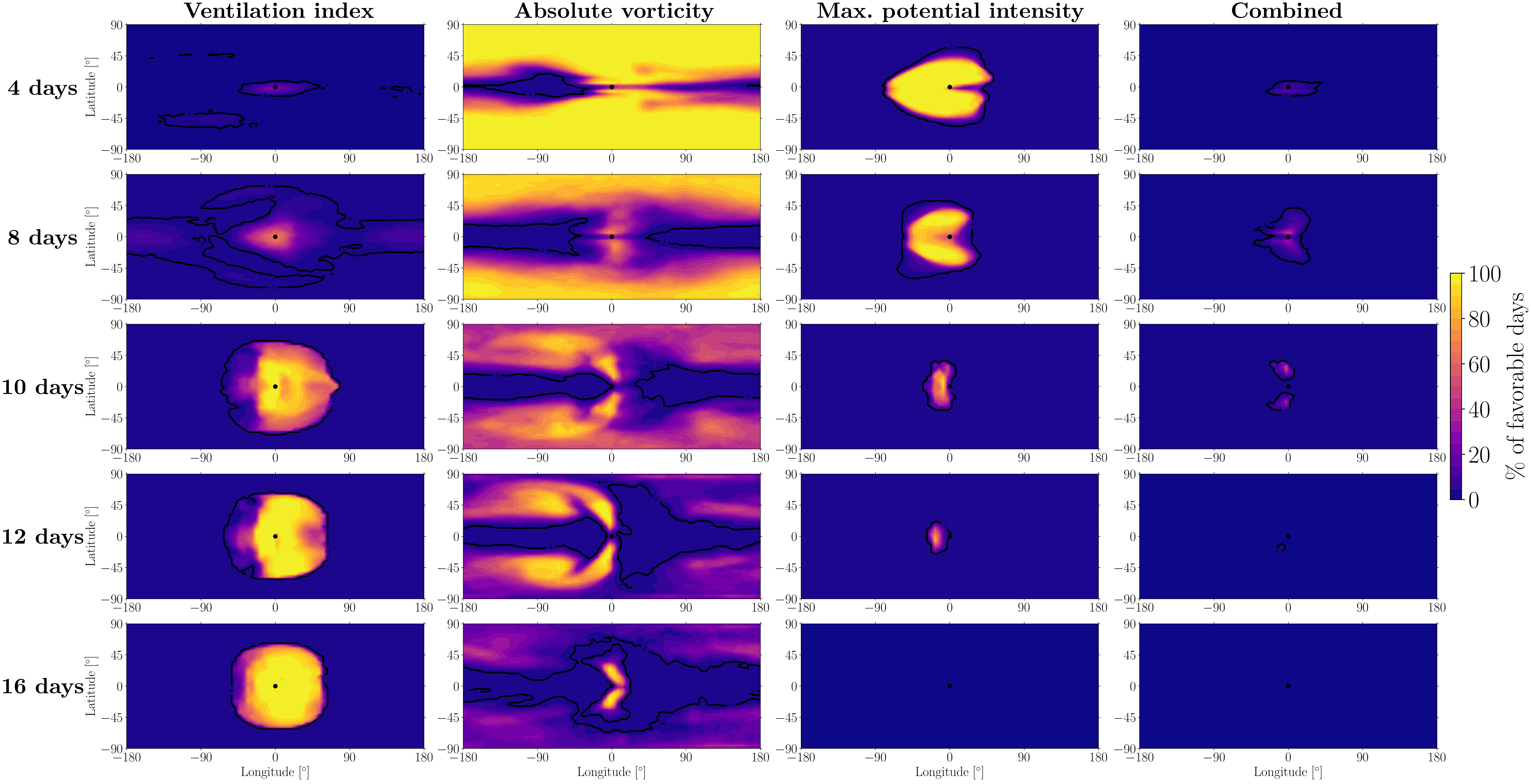}
\caption{
\textbf{Hurricane genesis is most favorable for tidally locked terrestrial exoplanets orbiting M dwarf stars with intermediate rotation periods.} Shown are maps of the percent of days that the hurricane genesis threshold is met in the environmental favorability metrics considered in this work, from left to right: ventilation index, lower-atmosphere absolute vorticity, maximum potential intensity, and all three combined. Contours display regions where hurricane genesis is favorable at least $1\%$ of the time. The dot at the center of each map indicates the substellar point. We show results from simulations with varying rotation rate alone, from 4 days to 16 days. We find that the maximum potential intensity and absolute vorticity thresholds are more often met for faster rotating planets, but the ventilation index threshold is met less often on fast rotating planets. As a result, all three metrics combined are most often satisfied for planets with intermediate rotation periods of $8-10~\mathrm{days}$.}
\label{fig:favorability_rot}
\end{sidewaysfigure*}
\indent We first analyze how environmental favorability depends on rotation period in order to investigate how the favorability of hurricane genesis relates to the dynamical regime of the atmosphere. \Fig{fig:favorability_rot} shows maps of the percent of time that the ventilation index, absolute vorticity, maximum potential intensity, and all three thresholds combined are met as a function of rotation period from 4 - 16 days. 
Generally, we find that hurricane genesis is most favorable in regions on the dayside north and south of the substellar point, which agrees with the results of \cite{Bin:2017aa} and \cite{Yang:2020}. We find that the ventilation index near the substellar point becomes more favorable with increasing rotation period. This is because of a reduction in the vertical wind shear with increasing rotation period as the planet moves from a fast-rotating to slow-rotating dynamical regime, with a corresponding reduction in the strength of the equatorial super-rotating jet \citep{Merlis:2010,Haqq2018}. However, both the absolute vorticity and maximum potential intensity become more favorable with \textit{decreasing} rotation period. The increase in absolute vorticity with decreasing rotation period is because faster-rotating planets have a larger absolute vorticity due to an increase in the planetary vorticity ($f$). The maximum potential intensity increases with decreasing rotation period because the vertical temperature contrast between the surface and tropopause, which sets the efficiency of the hurricane heat engine, is larger for planets with slower rotation periods. \\
\indent The opposing trends with rotation period in the ventilation index compared to the absolute vorticity and maximum potential intensity shown in \Fig{fig:favorability_rot} imply that environmental favorability for hurricane genesis peaks at intermediate rotation periods. When varying rotation period alone, simulations with rotation periods between $8~\mathrm{days}$ and $10~\mathrm{days}$ are most likely to have favorable regions for hurricane genesis. Faster rotating planets have enhanced mid-level ventilation, which reduces the favorability of hurricane genesis. Additionally, more slowly rotating planets have a maximum potential intensity below the hurricane threshold. 
As a result, we find that slowly rotating planets orbiting M dwarf stars likely do not host hurricanes, but may still experience weaker systems such as tropical storms and depressions. \\
\begin{table}
\begin{center}
\resizebox{0.45\textwidth}{!}{%
\begin{tabular}{| c | c |} 
\hline
{\bf Simulation parameters} & {\bf Favorability ($\%$)}  \\
\hline
\uline{Rotation period} &   \\
4 days & 47.5 \\
8 days & 91.1 \\
10 days & 83.3 \\
12 days & 6.9 \\
16 days & 0.0 \\
\hline
\uline{Planetary radius} &   \\
\textit{8 day rotation period:} &  \\
0.5 $R_\varoplus$ & 0.0  \\
2 $R_\varoplus$ & 0.0 \\
\textit{16 day rotation period:} &  \\
0.5 $R_\varoplus$ & 0.0 \\
2 $R_\varoplus$ & 33.3 \\
\hline
\uline{Surface gravity}  & \\
\textit{8 day rotation period:} &  \\
0.707 $g_\varoplus$ & 93.1 \\
1.414 $g_\varoplus$ & 96.1 \\
\textit{16 day rotation period:} &  \\
0.707 $g_\varoplus$ & 0.0 \\
1.414 $g_\varoplus$ & 0.0 \\
\hline
\uline{Surface pressure} &   \\
\textit{8 day rotation period:} &  \\
0.5 bars & 100  \\
4 bars & 6.4  \\
\textit{16 day rotation period:} &  \\
0.5 bars & 87.8  \\
4 bars & 8.3 \\
\hline
\uline{Incident stellar flux and rotation period} &   \\
\uline{(stellar $T_\mathrm{eff} = 2600~\mathrm{K}$)} & \\
0.882 $F_\varoplus$, 4.51 days & 96.4  \\
0.919 $F_\varoplus$, 4.37 days & 98.6  \\
0.955 $F_\varoplus$, 4.25 days & 98.6 \\
0.992 $F_\varoplus$, 4.13 days & 55.3 \\
1.011 $F_\varoplus$, 4.07 days & 1.1 \\
\hline
\uline{Incident stellar flux and rotation period} &  \\ 
\uline{(stellar $T_\mathrm{eff} = 3000~\mathrm{K}$)} & \\
0.955 $F_\varoplus$, 8.83 days & 84.2  \\
1.029 $F_\varoplus$, 8.35 days & 91.7  \\
1.103 $F_\varoplus$, 7.93 days & 88.3   \\
1.139 $F_\varoplus$, 7.74 days & 5.3  \\
\hline
\uline{Incident stellar flux and rotation period} &  \\ 
\uline{(stellar $T_\mathrm{eff} = 3300~\mathrm{K}$)} & \\
0.816 $F_\varoplus$, 26.4 days &  1.4  \\
1 $F_\varoplus$, 22.7 days &  0.0 \\
\hline
\uline{Incident stellar flux and rotation period} &  \\ 
\uline{(stellar $T_\mathrm{eff} = 4000~\mathrm{K}$)} & \\
0.816 $F_\varoplus$, 86.6 days & 23.6 \\
1 $F_\varoplus$, 74.3 days & 24.4  \\
\hline
\uline{Sensitivity tests} & \\ 
\uline{(8 day rotation period)} & \\ 
Inc. CO$_2$, CH$_4$ & 87.7 \\
$1~\mathrm{m}$ ocean & 82.2 \\
\hline
\end{tabular}}
\caption{\textbf{Hurricane genesis is favorable on a wide range of tidally locked terrestrial exoplanets.} Shown is the percent of days that hurricane genesis is favorable from GCM simulations for planets orbiting late-type stars (the host star has $T_\mathrm{eff} = 2600~\mathrm{K}$ unless otherwise marked). 
We find that hurricane genesis is most favorable on planets with intermediate rotation periods, and can occur over a wide range of incident stellar flux, surface gravity, and surface pressure. Note that the top row of simulations with varying rotation period correspond to the same simulations shown in \Fig{fig:favorability_rot}.}
\label{table:favorability}
\end{center}
\end{table}
Results for the environmental favorability of hurricane genesis from our full suite of GCM simulations are shown in \Tab{table:favorability}. These favorability values are larger than those in \Fig{fig:favorability_rot} because now we analyze whether hurricane genesis is favorable at any location on the planet. We find that at a fixed incident stellar flux equal to that of Earth, hurricane genesis is favorable most of the time on Earth-sized planets with intermediate rotation periods between $8~\mathrm{days}$ and $10~\mathrm{days}$. However, hurricane genesis is not always favorable, with a peak at $91.1\%$ for the 8 day rotation period case. This is because the ventilation index is most favorable near the substellar point while the maximum potential intensity peaks at mid-latitudes, and weather causes the regions of favorable ventilation index and maximum potential intensity to occasionally not overlap. As discussed above, we find that both faster and slower rotating planets are less favorable for hurricane genesis than intermediate rotators. Notably, we find that there is significantly reduced favorability for hurricane genesis in slowly rotating simulations of Earth-sized planets that have surface pressures equal to that of Earth and rotation periods $\ge 16~\mathrm{days}$. \\
\indent \Tab{table:favorability} also shows how the favorability of hurricane genesis depends on planetary radius, surface gravity, and surface pressure for planets with an intermediate rotation period of $8~\mathrm{days}$ and a slow rotation period of $16~\mathrm{days}$. We find that environmental favorability is sensitive to planetary parameters, with planets that were favorable for hurricane genesis with Earth-like parameters becoming less favorable with varying radius and higher surface pressure. We also find that the maximum potential intensity does not reach the hurricane threshold when varying gravity in the slowly rotating regime. Interestingly, we find that the maximum potential intensity can increase such that slowly rotating planets with lower surface pressures than Earth are favorable for hurricane genesis. \\
\indent For a fixed rotation period of 8 days, we find that increasing the radius to $2~R_\varoplus$ and increasing the surface pressure to $4~\mathrm{bars}$ causes hurricane genesis to become less favorable. Both cases have ventilation indices near the substellar point that are too large to consistently allow for hurricane genesis. Increasing radius has a similar dynamical effect to increasing the rotation rate \citep{Chavas:2019aa,Yang:2019aa}, reducing the substellar cloud cover and causing the planet to reach a hot climate state. Note that we find that hurricane genesis is also unfavorable in simulations with a reduced radius of $0.5~R_\varoplus$, as the maximum potential intensity is below the hurricane threshold. Increased surface pressure also warms the surface due to the increased column mass of water vapor \citep{Kopp:2014,Yang:2019aa}, leading to stronger vertical wind shear, which enhances the ventilation index. The effect of increasing surface pressure causes hurricane genesis to become significantly less favorable in the 8 day rotation period case but does not greatly affect favorability in the 16 day period case. This is because the ventilation index near the substellar point is already near the threshold value in the 8 day rotation period case, which amplifies the effect of small changes in ventilation. 
\subsubsection{Dependence on host star type}
\indent To determine how environmental favorability depends on host star type, we varied the incident stellar flux and rotation period self-consistently for planets orbiting M dwarf host stars with effective temperatures ranging from $2600~\mathrm{K}$ to $4000~\mathrm{K}$ (\Tab{table:favorability}). We find that hurricane genesis can be favorable in our simulations of planets orbiting late-type M dwarfs with effective temperatures of $2600~\mathrm{K}$ and $3000~\mathrm{K}$. For planets orbiting late-type M dwarfs, we find that hurricane genesis is generally favorable for planets with an incident stellar flux that is comparable to or smaller than that of Earth. This is because the vertical wind shear is reduced in cool climates due to the weakened strength of atmospheric superrotation relative to simulations closer to the inner edge of the habitable zone. Because of strong atmospheric superrotation, we find that hurricane genesis is significantly less favorable for planets that orbit near the inner edge of the habitable zones of late-type M dwarf stars. The reduced favorability for hurricane genesis near the inner edge of the habitable zone of late-type M dwarf stars is due to strong vertical wind shear, which overwhelms the increase in maximum potential intensity and leads to larger ventilation indices. However, the exact location of the inner edge of the habitable zone for a given stellar type may be affected by the choice of GCM \citep{Kopp:2016,Bin:2018aa,YangJun:2019}. Regardless, the qualitative result of decreased favorability for hurricane genesis near the inner edge should apply regardless of the exact value of incident stellar flux at which the inner edge of the habitable zone lies. \\
\indent We find that, at an incident stellar flux of $\sim 1 F_\varoplus$, the favorability of hurricane genesis is much smaller in our simulations of planets that orbit stars with effective temperatures of $3300~\mathrm{K}$ and $4000~\mathrm{K}$ than for planets orbiting late-type M dwarfs. This is because the slow rotation of tidally locked planets orbiting earlier-type stars leads to smaller maximum potential intensities, which reduces the favorability of hurricane genesis. However, the ventilation index is generally favorable for tropical cyclogenesis in all of our simulations of planets orbiting stars with effective temperatures of $3300~\mathrm{K}$ and $4000~\mathrm{K}$. As a result, we find that tidally locked planets orbiting earlier-type red dwarf stars that receive an incident stellar flux similar to or slightly lower than that of Earth may still be favorable for the genesis of weaker tropical storms. However, our simulations do not probe near the inner edge of the habitable zone of early-type M dwarf stars. Planets near the inner edge of the habitable zone of early-type M dwarf stars could be favorable for hurricane genesis, as they are in a slowly rotating dynamical regime \citep{Haqq2018} and have reduced vertical wind shear relative to planets near the inner edge of the habitable zone of late-type M dwarf stars.
\subsubsection{Sensitivity tests}
\indent To test the sensitivity of our results to assumptions in our model setup, we ran two additional simulations with a rotation period of 8 days: one with a shallower slab ocean depth (reduced from $50~\mathrm{m}$ to $1~\mathrm{m}$) and a second with modern Earth-like abundances of $400~\mathrm{ppm}$ CO$_2$ and $1.7~\mathrm{ppm}$ CH$_4$ rather than considering a pure N$_2$-H$_2$O atmosphere (\Tab{table:favorability}). We find that both cases have slightly reduced favorability, but hurricane genesis is still favorable $87.7\%$ of the time in the simulations including greenhouse gases and $82.2\%$ of the time in the simulation with a $1~\mathrm{m}$ deep slab ocean. As a result, changing our assumptions of the greenhouse gas composition and slab ocean depth does not greatly impact hurricane favorability. We discuss the mean climate from these sensitivity tests in more detail in Appendix \ref{app:sens}. 
\subsection{Ventilation-reduced maximum potential intensity}
\begin{figure}
\centering
\includegraphics[width=0.48\textwidth]{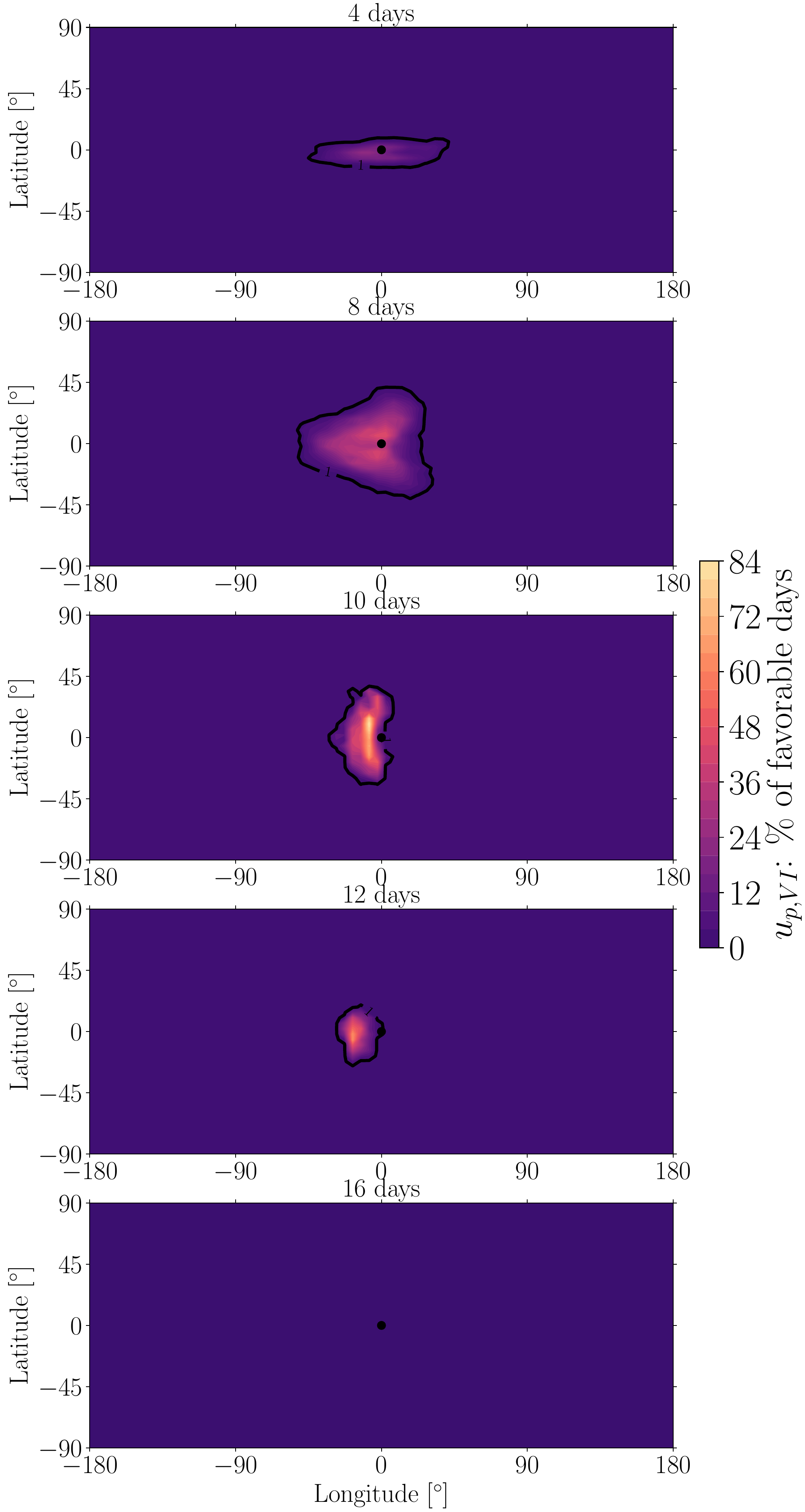}
\caption{\textbf{The ventilation-reduced maximum potential intensity captures the energetic effect of importing dry, low-entropy air on the maximum achievable wind speeds.} Colors show how often the ventilation-reduced maximum potential intensity exceeds the $33~\mathrm{m}~\mathrm{s}^{-1}$ threshold for hurricane genesis from simulations with varying rotation periods from 4 to 16 days. Contours show where the ventilation-reduced maximum potential intensity is favorable for hurricane genesis at least $1\%$ of the time. 
We find that the trend in ventilation-reduced maximum potential intensity differs from that of maximum potential intensity. The maximum potential intensity increases with decreasing rotation period, but the ventilation-reduced maximum potential intensity is most favorable for hurricane genesis in simulations with intermediate rotation periods of $\sim 10~\mathrm{days}$, as was found in the combined favorability metric of \Fig{fig:favorability_rot} (right-most column). This is because strong ventilation at short rotation periods causes the ventilation-reduced maximum potential intensity to decrease.}
\label{fig:vpvi}
\end{figure}
In this section, we determine if the ventilation-reduced maximum potential intensity can provide one metric that encapsulates our results for environmental favorability from \Fig{fig:favorability_rot}. In \Fig{fig:vpvi} we show how the percent of time that the ventilation-reduced maximum potential intensity $u_{p,\mathrm{VI}}$ is favorable for hurricane genesis depends on rotation period, where $u_{p,\mathrm{VI}}$ is calculated as described in \Sec{sec:othermetrics}. 
Similarly to our combined hurricane favorability metrics, we find that the ventilation-reduced maximum potential intensity is large enough near the substellar point that  hurricane genesis is commonly favorable for intermediate rotation periods of $\sim 10~\mathrm{days}$. The favorability of the ventilation-reduced maximum potential intensity also similarly decreases toward both high and low rotation periods. As a result, we find that the ventilation-reduced maximum potential intensity provides a single metric that matches the broad trends we find in our full suite of combined metrics. \\
\indent The behavior of the ventilation-reduced maximum potential intensity at short rotation periods is the opposite of the maximum potential intensity itself (shown in the rightmost column of \Fig{fig:favorability_rot}). At short rotation periods, the maximum potential intensity increases, whereas the ventilation-reduced maximum potential intensity decreases due to the enhanced effect of ventilation and resulting unfavorable ventilation index (shown in the leftmost column of \Fig{fig:favorability_rot}). At long rotation periods $> 10$ days, the favorability of the ventilation-reduced maximum potential intensity is similar to that of the maximum potential intensity because the ventilation index near the substellar point is largely below the threshold at which ventilation disrupts hurricane genesis.  
\section{Observational implications of hurricane genesis}
\label{sec:disc}
Because hurricanes affect global-mean climate and enhance vertical transport of water vapor, they may impact future observations of terrestrial exoplanets. \cite{Yang:2020} proposed that hurricanes near the terminator can transport water vapor to low pressures, causing the transmission spectral signature of water vapor to change with time. When hurricanes are near the terminator, \cite{Yang:2020} found that the local specific humidity can be enhanced by up to a factor of $\sim 5$. Next-generation infrared space-based telescopes (e.g., OST) may provide water vapor abundance constraints to within an order of magnitude \citep{Tremblay:2020aa}, and down to a factor of $\sim 5$ with up to 100 transits. As a result, hurricanes may cause detectable changes in the water vapor abundance at the terminator -- however, note that cloud cover may affect the detectability of water vapor features in transmission \citep{Fauchez:2019aa,Suissa:2019aa,Komacek:2020aa}.  \\
\indent Tidally locked terrestrial exoplanets orbiting M dwarf stars are expected to have continuous cloud cover at the substellar point due to dayside convection \citep{Yang:2013,way:2018}. It is possible that hurricanes will cause this cloud cover to be time-variable, due to the enhanced vertical transport of water vapor. Cloud systems on Earth, including hurricanes, readily self organize \citep{Holloway:2017aa}, and hurricanes have fast intrinsic dynamical timescales \citep{Emanuel:2012aa}. This suggests that if hurricanes form, they could be a dominant type of convective aggregation. Thus, hurricanes may focus convective clouds in a small area while suppressing clouds and drying out the atmosphere in the broader environment \citep{Holloway:2017aa}, especially in moist climates \citep{Chavas:2015aa,Cronin:2019aa}. 
Cloud-resolving models of terrestrial exoplanets have shown that the cloud cover on the dayside can be strongly spatially and/or temporally variable due to convective aggregation (\citealp{Sergeev:2020aa}, D.D.B. Koll, personal comm.). Hurricanes have been shown to cause significant variability in the temporal and spatial distribution of clouds and precipitation on terrestrial exoplanets \citep{Yang:2020}, and may thus also enhance the variability in dayside cloud cover. Variable dayside cloud cover may affect the shape of the planetary phase curve \citep{Yang:2014,Haqq2018}, which could be probed through future thermal emission or reflected light observations. \\
\indent This work has important implications for both exoplanets and Earth. Given that existing hurricane theory has been validated against Earth data out of necessity, Earth-based theories serve as our only current source of knowledge for understanding how hurricanes may behave on other planets. Our numerical simulations may be tested with future observations that constrain how hurricanes impact vapor transport and cloud formation in exoplanet atmospheres. Our results also provide a foundation for further testing using high-resolution models that explicitly resolve hurricanes. Meanwhile, an ideal means of evaluating our basic understanding of hurricane activity on Earth is to test our Earth-based theories on planetary climates wildly different from Earth. This study provides insights into the application of Earth-based theories to tidally locked terrestrial planets orbiting late-type stars. More generally, this work broadens the horizons for how Earth-based theories may be used to understand weather on terrestrial exoplanets.
\section{Conclusions}
\label{sec:conc}
In this work, we applied a combination of Earth-based metrics for the favorability of hurricane genesis to three-dimensional climate simulations of tidally locked terrestrial exoplanets orbiting M dwarf stars. Overall, we find that hurricane genesis may be favorable on certain exoplanets orbiting late-type M dwarf stars. Our conclusions are as follows:  
\begin{enumerate}
\item Hurricane genesis is favorable on terrestrial exoplanets orbiting M dwarf stars with intermediate rotation periods. This is because of the competing effects of increasing absolute vorticity with decreasing rotation period, which enhances the favorability of hurricane genesis, and increasing ventilation and wind shear with decreasing rotation period, which decreases favorability. As a result, we find that hurricane genesis is most favorable on tidally locked terrestrial exoplanets with intermediate rotation periods of $\sim 8-10~\mathrm{days}$. Using additional simulations with consistently varying incident stellar flux and rotation period, we find that hurricane genesis is most favorable on terrestrial exoplanets in the habitable zones of late-type M dwarf stars with effective temperatures $T_\mathrm{eff} \lesssim 3000~\mathrm{K}$. Genesis favorability changes only modestly with the inclusion of additional Earth-like abundances of carbon dioxide and methane as well as the use of a much shallower slab ocean.
\item We find that hurricane genesis is not favorable on slowly rotating tidally locked terrestrial planets. This result agrees with the previous work of \cite{Bin:2017aa}, who found that hurricanes are unlikely to occur in simulations of planets with a rotation period of 28 days around a star with $T_\mathrm{eff} = 3700~\mathrm{K}$. Additionally, we estimate that for Earth-sized planets that receive the same incident stellar flux as Earth and have surface pressures equal to that of Earth, the threshold rotation period above which hurricane genesis is unfavorable is $\gtrsim 16~\mathrm{days}$. As a result, our simulations suggest that hurricane genesis is less likely to occur on tidally locked exoplanets that receive a comparable amount of incident flux to Earth and orbit early-type M dwarf stars. 
\end{enumerate}
\acknowledgements
We thank Eric Wolf for making \exocam~freely available to the community. We thank the referee for thoughtful comments that improved this work. We acknowledge support from the NASA Astrobiology Institute Virtual Planetary Laboratory, which is supported by NASA under cooperative agreement NNH05ZDA001C. This work was partially supported by the NASA Astrobiology Program Grant Number 80NSSC18K0829 and benefited from participation in the NASA Nexus for Exoplanet Systems Science research coordination network. T.D.K. acknowledges funding from the 51 Pegasi b Fellowship in Planetary Astronomy sponsored by the Heising-Simons Foundation. Our work was completed with resources provided by the University of Chicago Research Computing Center. 
\appendix
\section{Mean favorability metrics}
\label{sec:appa}
\begin{sidewaysfigure*}
\centering
\vspace{-3in}
\includegraphics[width=1\textwidth]{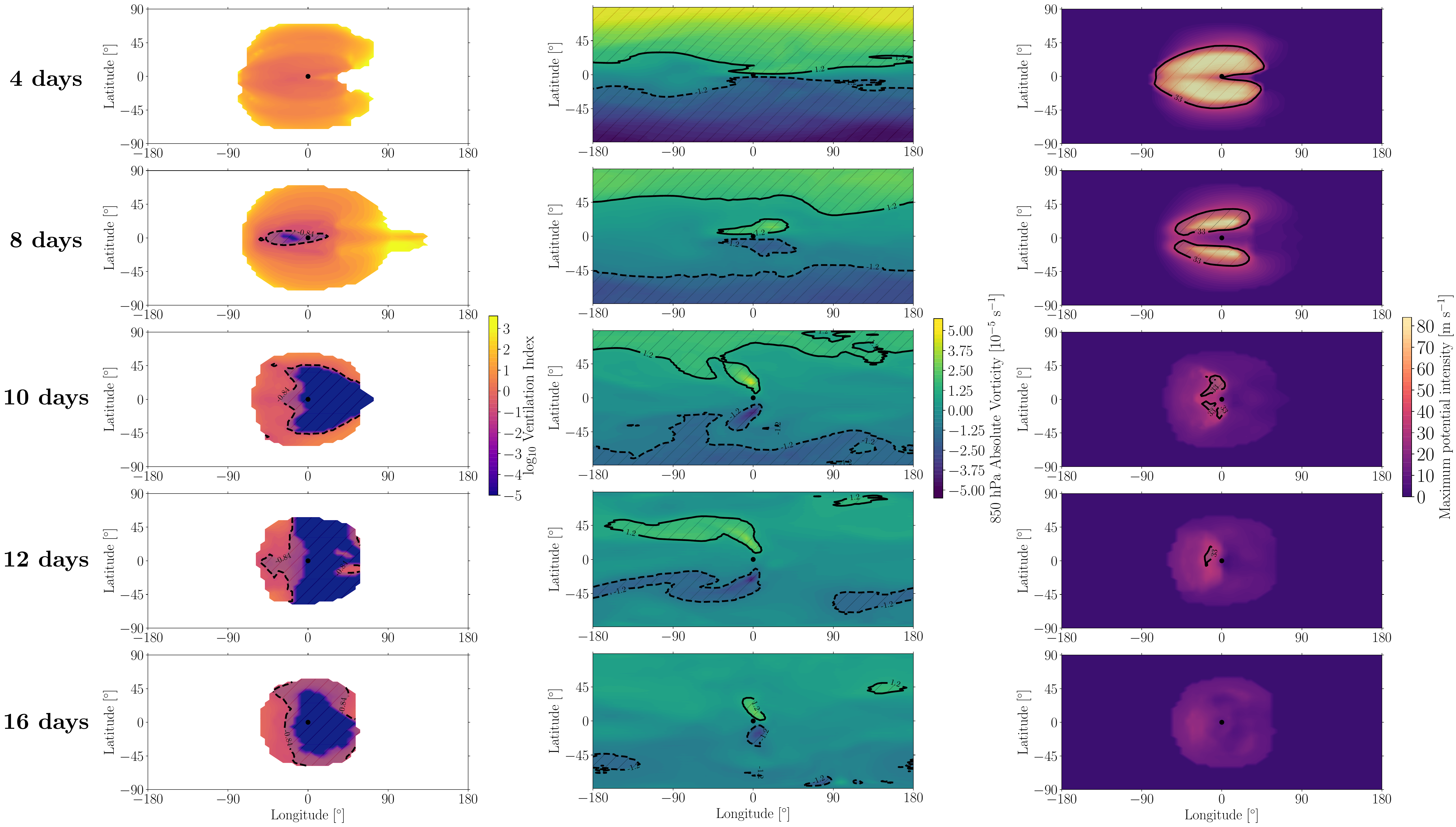}
\caption{
\textbf{Results for the environmental favorability of hurricane genesis from mean climate metrics are similar to our analysis using daily output.} Colors show the monthly mean of the ventilation index (left column), absolute vorticity (center column), and maximum potential intensity (right column), while the hatched regions surrounded by contours show regions that are favorable for hurricane genesis from each metric. We show results from the same simulations in \Fig{fig:favorability_rot} with varying rotation rate alone, from 4 days to 16 days. We find the same trend in increasing favorability in absolute vorticity and maximum potential intensity but decreasing favorability in the ventilation index with decreasing rotation period found in our nominal analysis, leading to a peak in the favorability of hurricane genesis at intermediate rotation periods.}
\label{fig:favorability_rot_mean}
\end{sidewaysfigure*}
\indent In this Appendix we present an alternate analysis of the hurricane genesis favorability in the mean climate state for a subset of simulations. We do so in order to compare with our main analysis in which we determine the percent of time that hurricane genesis is favorable. 
\subsection{Varying rotation period}
\Fig{fig:favorability_rot_mean} shows monthly averages of the ventilation index, lower-atmosphere absolute vorticity, and maximum potential intensity from simulations with varying rotation period alone from 4 - 16 days. This is analogous to \Fig{fig:favorability_rot}, but showing the hurricane genesis favorability in the mean climate rather than the percent of time that hurricane genesis is favorable.  We find that our results from the mean climate are similar to our analysis from daily output. The ventilation index increases (becomes less favorable) with decreasing rotation period, while both the absolute vorticity and maximum potential intensity increase (become more favorable) with decreasing rotation period. As a result, we expect from the mean climate that hurricane genesis is most favorable at intermediate rotation periods of $\sim 10~\mathrm{days}$, as we found in our main analysis. 
\subsection{Sensitivity tests}
\label{app:sens}
\begin{sidewaysfigure*}
\centering
\vspace{-3in}
\includegraphics[width=1\textwidth]{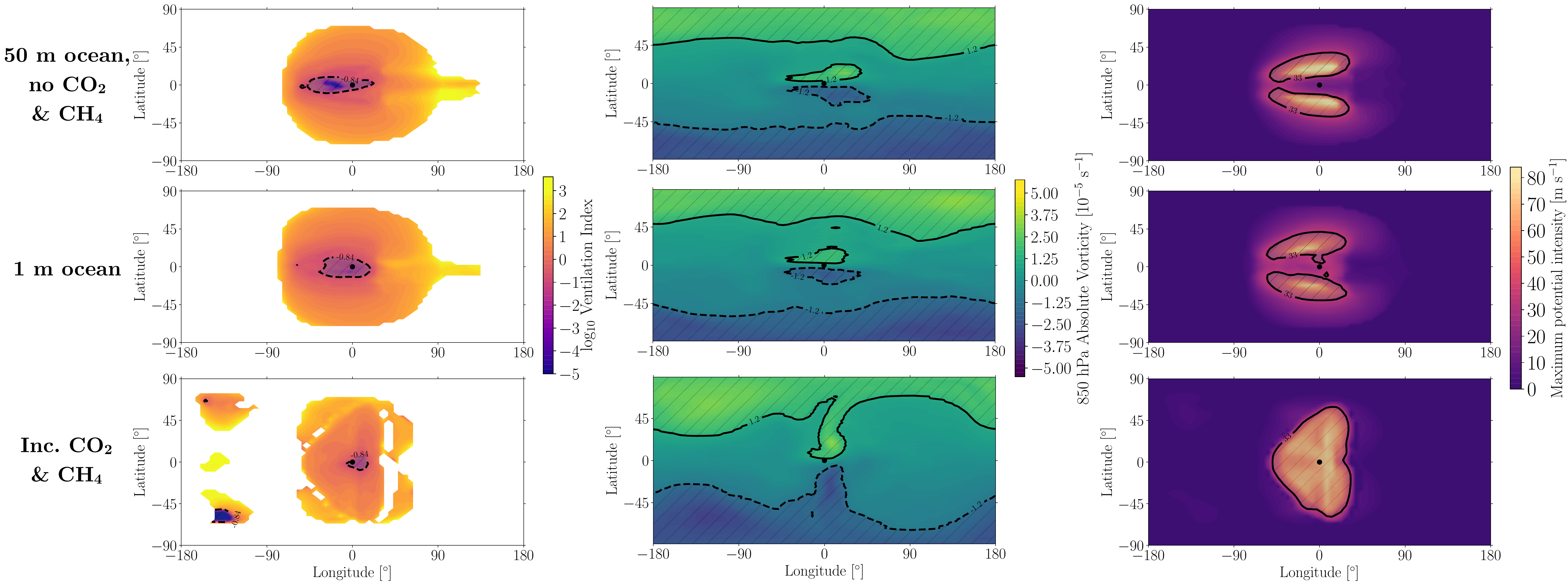}
\caption{\textbf{Our results for the environmental favorability of hurricane genesis are robust to assumptions for the ocean depth and atmospheric composition.} Shown are the mean ventilation index (left column), absolute vorticity (center column), and maximum potential intensity (right column) metrics from three GCM simulations. The simulation in the top row is our standard $8~\mathrm{day}$ rotation case, in which hurricanes are favorable in a narrow strip near the substellar point. The simulation in the middle row uses a shallower ocean of 1 meter depth but with all other parameters the same as the case in the top row, and the simulation in the bottom row includes modern Earth-like abundances of CO$_2$ and CH$_4$ with all other parameters the same as the case in the top row. Hatched regions surrounded by contours show where hurricane genesis is favorable from each metric. We find that although varying the slab ocean depth and atmospheric composition does affect climate, hurricane genesis is favorable in the mean climate state for all three cases.}
\label{fig:sensitivity}
\end{sidewaysfigure*}
\Fig{fig:sensitivity} shows favorability metrics calculated from the mean climate for our sensitivity tests varying the greenhouse gas concentration and slab ocean depth. We find that reducing the slab ocean depth did not significantly affect our results, as hurricane genesis remains favorable at shallow slab ocean depth. We also find that including Earth-like abundances of greenhouse gases increases the wind speed and maximum potential intensity near the substellar point. This increased near-surface wind speed leads to stronger vertical wind shear which acts to enhance the ventilation near the substellar point. However, the ventilation index generally still lies below the hurricane genesis threshold, and we find that hurricane genesis remains favorable when including Earth-like abundances of CO$_2$ and CH$_4$. These results from the mean climate are similar to those found in our standard favorability analysis from daily output shown in \Tab{table:favorability}. 
\subsection{Comparison to genesis potential index}
\label{app:gpi}
\begin{figure}
\centering
\includegraphics[width=0.45\textwidth]{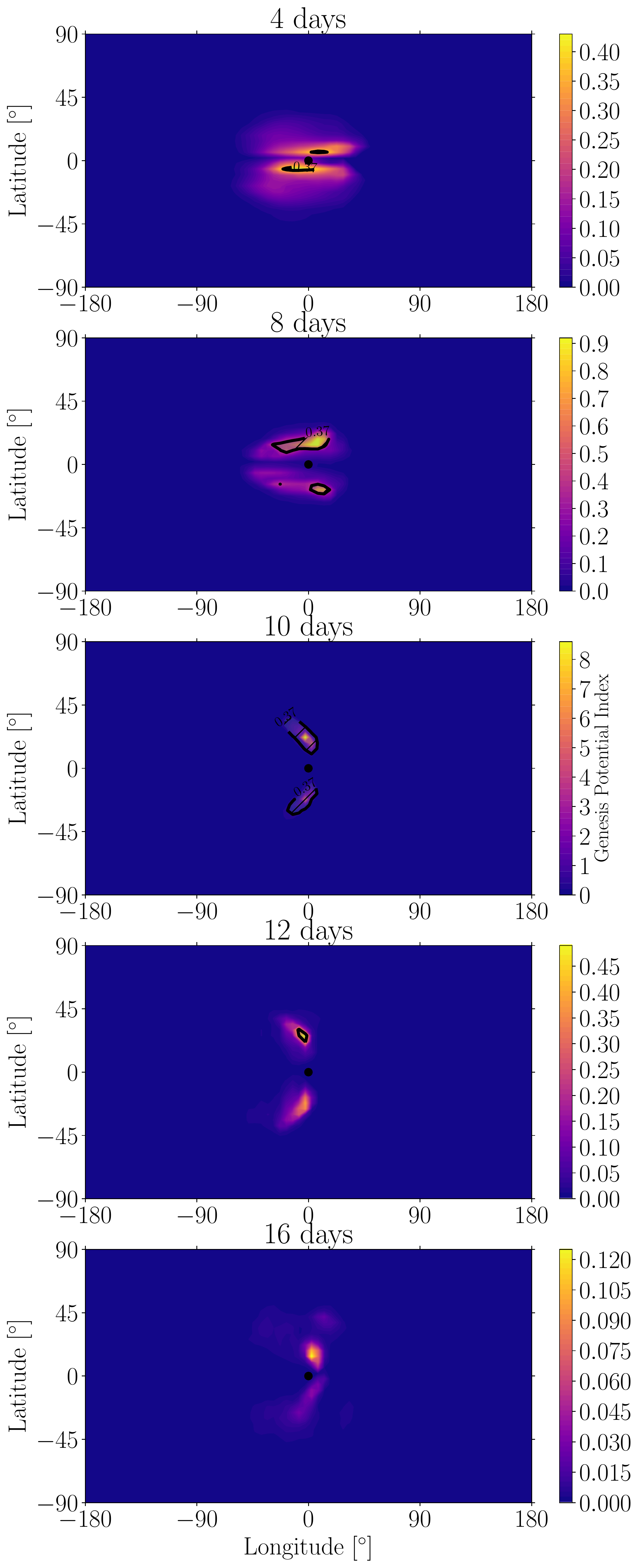}
\caption{\textbf{The favorability of hurricane genesis using the genesis potential index generally agrees with the favorability calculated from our combined metrics.} Shown are maps of the genesis potential index calculated from the mean climate of simulations with varying rotation periods from 4 to 16 days. Contours show where the hurricane genesis threshold of $\mathrm{GPI} = 0.37$ used in \cite{Bin:2017aa} is satisfied.
The genesis potential index peaks at a rotation period of 10 days, in agreement with the environmental favorability determined from our combined metrics. The decreasing genesis potential index with increasing rotation period above 10 days agrees with \cite{Bin:2017aa}, who found that hurricane genesis is not favorable on slowly rotating tidally locked terrestrial exoplanets.}
\label{fig:GPI}
\end{figure}
\indent To compare our results to previous work that determined the favorability of hurricane genesis on tidally locked exoplanets, we also computed the genesis potential index (Equation \ref{eq:GPI}) from our suite of GCM simulations. \Fig{fig:GPI} shows maps of the genesis potential index from the same simulations as shown in \Fig{fig:favorability_rot}.
Similar to our combined favorability metrics, the genesis potential index peaks for intermediate rotators with rotation periods of $\sim 10~\mathrm{days}$ and falls off toward faster and slower rotation. This is because intermediate rotators have low vertical wind shear but sufficient vorticity and maximum potential intensity, all of which increase the genesis potential index. Our results for the genesis potential index agree with the analysis of \cite{Bin:2017aa}, as the genesis potential index is low in our simulations of slowly rotating tidally locked planets with circular orbits, which aligns with their finding that hurricane genesis is unlikely to occur on planets orbiting M dwarf stars that have a rotation period of 28 days. We computed the genesis potential index from our entire suite of simulations, and find general agreement with our combined metrics.

\end{document}